\documentclass[pra,showpacs]{revtex4}
\usepackage{makeidx}
\usepackage{amsmath}
\usepackage{graphicx}
\usepackage{epsfig}
\usepackage[usenames]{color}
\usepackage{subfigure}

\newcommand{\beq}{\begin{equation}}
\newcommand{\eeq}{\end{equation}}
\newcommand{\beqa}{\begin{eqnarray}}
\newcommand{\eeqa}{\end{eqnarray}}
\newcommand{\ba}{\begin{array}}
\newcommand{\ea}{\end{array}}

\begin{document}

\title{Solitons and solitary vortices in ``pancake"-shaped Bose-Einstein
condensates}
\author{Luca Salasnich$^1$ and Boris A. Malomed$^2$}
\affiliation{$^{1}$CNR-INFM and CNISM, Unit\`a di Padova, Dipartimento di Fisica
``Galileo Galilei'', Universit\`a di Padova, Via Marzolo 8, 35131 Padova,
Italy \\
$^{2}$Department of Physical Electronics, School of Electrical Engineering,
Faculty of Engineering, Tel Aviv University, Tel Aviv 69978, Israel}

\begin{abstract}
We study fundamental and vortical solitons in disk-morphed
Bose-Einstein condensates (BECs) subject to strong confinement along
the axial direction. Starting from the three-dimensional (3D)
Gross-Pitaevskii equation (GPE), we proceed to an effective 2D
nonpolynomial Schr\"{o}dinger equation (NPSE) derived by means of
the integration over the axial coordinate. Results produced by the
latter equation are in very good agreement with those obtained from
the full 3D GPE, including cases when the formal 2D equation with
the cubic nonlinearity is unreliable. The 2D NPSE is used to predict
the density profiles and dynamical stability of repulsive and
attractive BECs with zero and finite topological charge in various
planar trapping configurations, including the axisymmetric harmonic
confinement and 1D periodic potential. In particular, we find a
stable dynamical regime that was not reported before, \textit{viz}.,
periodic splitting and recombination of trapped vortices with
topological charge $2$ or $3$ in the self-attractive BEC.
\end{abstract}

\maketitle

\section{Introduction}

A natural setting for the creation of localized states, both fundamental
ones and those carrying intrinsic vorticity, in Bose-Einstein condensates
(BECs) is provided by ``pancake" configurations, which are strongly confined
in one (axial) direction, $z$, being weakly trapped in the transverse plane,
($x,y$). In the experiment, ``pancakes" were created by a superposition of a
tight flat optical trap, formed by a pair of strongly repulsive
(blue-detuned) light sheets, and a loose in-plane radial magnetic trap \cite%
{Ketterle}. Alternatively, the axial trap may be represented by one cell of
a very strong optical lattice (OL)\ \cite{Hulet}.

The pancake configuration may exist with either sign of the intrinsic
nonlinearity, self-repulsive or self-attractive (although, as far as we
know, in the latter case experiments were not reported, as yet, in the
pancake geometry). In particular, the attractive sign can be induced by
means of the Feshbach-resonance technique, as in well-known works which
demonstrated the creation of effectively one-dimensional (1D) solitons in
the condensates of $^{7}$Li \cite{soliton} and $^{85}$Rb \cite{soliton2}
atoms, and in other settings (for instance, in experiments with the
condensate formed by weakly interacting $^{39}$K atoms \cite{Inguscio}).
Depending on the sign of the nonlinearity, one may expect the establishment
of different matter-wave patterns. In the case of the repulsion, a
relatively weak (in comparison with the tight axial trap) in-plane parabolic
(alias harmonic) confining potential gives rise to an axisymmetric ground
state, that may be described by means of the Thomas-Fermi (TF) approximation
\cite{BEC}. Imparting the angular moment to the confined axisymmetric state
creates a vortex, which may be unstable against the bending of its axis in
the full 3D geometry \cite{bending}, but is stable in the planar form. On
the other hand, multiple vortices, with topological charge $S\geq 2$, were
demonstrated experimentally \cite{double-vort-exper} and theoretically \cite%
{double-vort} to be unstable against splitting into unitary ones.
Nevertheless, it was predicted that multiple vortices may be stabilized by
imposing an anharmonic in-plane trapping potential \cite{multiple}.

In the case of the self-attractive nonlinearity, a challenging problem is to
predict, and create in an experiment, 2D solitons and solitary vortices
(solitons with the embedded vorticity) that would be stable in the pancake
setting, despite the possibility of the collapse. If the in-plane
confinement is provided by the axisymmetric parabolic potential,
soliton-like states are stable provided that their norm (the total number of
atoms) falls below a critical value which stipulates the onset of the
collapse \cite{Dodd}. The critical value strongly increases for vortex
solitons, but they are vulnerable, below the collapse threshold, to another
instability, which tends to break them into a few fragments shaped as
fundamental solitons, which may later suffer the intrinsic collapse.
Stability limits for the trapped vortices in the self-attractive BEC have
been studied in detail theoretically, both for the nearly planar and full 3D
geometries \cite{Sadhan}-\cite{DumDum2}, including the limit case of the
cigar-shaped (prolate) trap, opposite to that of the oblate ``pancake" \cite%
{Luca,we-vortex-in-tube}. In particular, it was found that such solitons
with multiple charges, $S\geq 2$, are always unstable; as concerns the
unitary vortices ($S=1$), they feature a region of a semi-stability between
fully stable and unstable configurations, where the vortex periodically
splits and recombines \cite{DumDum,DumDum2}.

Another general possibility to stabilize both fundamental solitons and
various solitary-vortex states in the 2D geometry is offered by the use of
periodic (rather than parabolic) in-plane potentials, that may be typically
created by means of OLs. For 2D periodic potentials, this possibility was
first demonstrated in Refs. \cite{BBB1} and \cite{Yang}. Then, it was shown
that a quasi-1D potential, induced by a 1D OL (i.e., a periodic potential
that depends on a single in-plane coordinate) can also stabilize fundamental
solitons \cite{BBB2}. The general case of an anisotropic potential, composed
by two perpendicular sublattices with unequal strengths, was investigated
recently \cite{Thawatchai}.

A very accurate model for the description of the mean-field dynamics of
dilute condensates is provided by the 3D Gross-Pitaevskii equation (GPE)
\cite{BEC}. An effective 2D equation for the ``pancake" geometry should be
derived from the 3D GPE by means of an adequate reduction procedure. It was
developed, using the ansatz which assumes the factorization of the wave
function in the axial direction and perpendicular (pancake's) plane, in Ref.
\cite{sala-npse} (see also Ref. \cite{Delgado}). The result is a
nonpolynomial nonlinear Schr\"{o}dinger equation(NPSE) (it takes a different
nonpolynomial form in the cigar-shaped setting, when the dimension is
reduced from 3 to 1 \cite{sala-npse}-\cite{we-twocomp}). The objective of
the present work is to use the 2D NPSE for the systematic analysis of
effectively 2D localized states in both cases of the self-repulsive and
self-attractive nonlinearity. To this end, we briefly recapitulate the
derivation of the effective 2D equation in Section II, where the respective
TF approximation is considered too. In Section III we focus on the model
with the repulsive nonlinearity. Using the 2D\ NPSE, we construct solution
families for both the ground state and vortices trapped in the axisymmetric
in-plane parabolic potential. The comparison with direct numerical solutions
of the underlying 3D GPE demonstrates that the effective equation, unlike
the formal 2D version of the GPE, with the cubic nonlinearity, provides for
virtually exact shapes of both the ground and vortex states. In Section IV,
we consider the model with the self-attraction. First, we construct families
of fundamental and vortical states trapped in the axisymmetric parabolic
potential. In that case too, we conclude that the respective version of the
2D NPSE provides for a virtually exact prediction of the shape of the
states. The 2D\ NPSE\ reproduces earlier known results for the stability of
trapped states with vorticity $S=1 $, and generates a new stable regime for $%
S=2$ and $3$, in the form of periodic splitting of the double or triple
vortex into a pair or triplet of unitary vortices followed by their
recombination back into the single double or triple vortex. Then, we
consider the quasi-1D periodic potential, for which families of fundamental
solitons are found as numerical solutions of the 2D NPSE, and also by means
of the variational approximation (VA) applied directly to the underlying 3D
GPE.

\section{The effective 2D equation}

\subsection{The general case}

We consider the condensate formed by a large number $N$ of bosonic atoms of
mass $m$ tightly trapped in axial direction $z$ by the harmonic potential, $%
(m/2)\omega _{z}^{2}z^{2}$, and loosely confined in transverse plane $(x,y)$
by a generic potential, $W(x,y)$. The scaled 3D GPE, which governs the
mean-field dynamics of the dilute potential at zero temperature, is
\begin{equation}
i{\frac{\partial }{\partial t}}\Psi =\left[ -{\frac{1}{2}}\nabla
^{2}+V(x,y,z)+2\pi g|\Psi |^{2}\right] \Psi \;,  \label{3dgpe}
\end{equation}%
where $V(x,y,z)\equiv W(x,y)+z^{2}/2$, $g\equiv 2Na_{s}/a_{z}$ is the
interaction strength, with $a_{s}$ the inter-atomic scattering length and $%
a_{z}=\sqrt{\hbar /(m\omega _{z})}$ the scale of the axial trapping, and 3D
wave function $\Psi (\mathbf{r},t)$ is subject to normalization
\begin{equation}
\int |\Psi (\mathbf{r},t)|^{2}\ d^{3}\mathbf{r}=1.  \label{norm}
\end{equation}

Following Ref. \cite{sala-npse}, the solution to Eq. (\ref{3dgpe}) can be
simplified by setting
\begin{equation}
\Psi (\mathbf{r},t)=\Phi (x,y,t)\ \Xi (z;\eta ),  \label{factor}
\end{equation}%
where $\Phi (x,y,t)$ is an arbitrary transverse wave function, and $\Xi
(z,\eta )$ is its normalized axial counterpart,
\begin{equation}
\Xi (z;\eta )=\left( \pi \eta ^{2}\right) ^{-1/4}\exp {\left[ -{z^{2}/}%
\left( 2\eta ^{2}\right) \right] },~\int_{-\infty }^{+\infty }\Xi
^{2}(z)dz=1.  \label{Xi}
\end{equation}%
With regard to underlying condition (\ref{norm}), the factorization ansatz
based on Eqs. (\ref{factor}) and (\ref{Xi}) yield the normalization
condition for the 2D wave function, $\int \int |\Phi (x,y,t)|^{2}\ dxdy=1$.
In these expressions, axial width $\eta $ may depend on transverse
coordinates $(x,y)$ and time $t$.

\begin{figure}[tbp]
{\includegraphics[width=8.cm,clip]{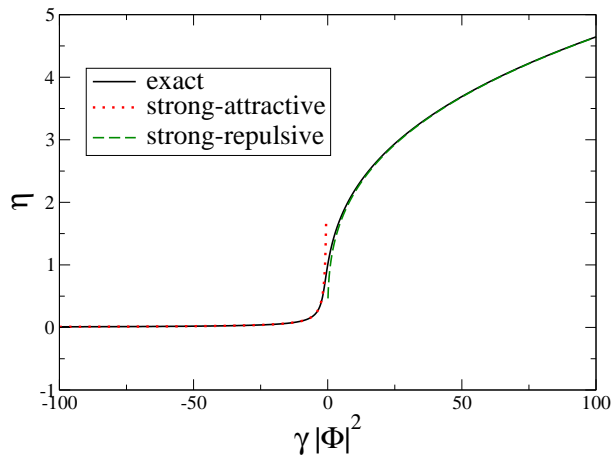}}
\caption{(Color online). Axial width $\protect\eta $ of the pancake-shaped
condensate as a function of the interaction-strength$\times $density
product, $\protect\gamma |\Phi |^{2}$. The solid curve: the full dependence
following from Eq. (\protect\ref{eta-solve}); dashed and dotted curves:
asymptotic expressions (\protect\ref{high}), in the cases of repulsion, $%
\protect\gamma >0$, and attraction, $\protect\gamma <0$ (except for the
narrow vicinity of $\protect\gamma |\Phi |^{2}=0$, the asymptotic curves are
virtually identical to the full one, in the regions of $\protect\gamma |\Phi
|^{2}>0$ and $\protect\gamma |\Phi |^{2}<0$, respectively).}
\label{fig_eta}
\end{figure}

The substitution of factorized expression (\ref{factor}) in Eq. (\ref{3dgpe}%
) and subsequent averaging in axial coordinate $z$ lead to the following 2D\
NPSE \cite{sala-npse},
\begin{equation}
i{\frac{\partial }{\partial t}}\Phi =\left[ -{\frac{1}{2}}\nabla _{\bot
}^{2}+W(x,y)+\gamma {\frac{|\Phi |^{2}}{\eta }}+{\frac{1}{4}}\left( {\frac{1%
}{\eta ^{2}}}+\eta ^{2}\right) \right] \Phi ,  \label{npse}
\end{equation}%
where the scaled interaction strength is%
\begin{equation}
\gamma =2\sqrt{2\pi }Na_{s}/a_{z}\equiv \sqrt{2\pi }g,  \label{pimpa}
\end{equation}%
and the axial width is determined by an algebraic equation,
\begin{equation}
\eta ^{4}=1+\gamma |\Phi |^{2}\eta .  \label{eta}
\end{equation}%
Exact solutions to Eq. (\ref{eta}) are given by the Cardano formula,
\begin{equation}
\eta =\pm {\frac{1}{2}}\sqrt{\frac{A^{2}-12}{3A}}+{\frac{1}{2}}\sqrt{-{\frac{%
A^{2}-12}{3A}}\pm 2\gamma |\Phi |^{2}\left( \frac{A^{2}-12}{3A}\right)
^{-1/2}}\;,  \label{eta-solve}
\end{equation}%
where the upper and lower signs correspond, respectively, to $\gamma >0$ and
$\gamma <0$, and
\begin{equation}
A\equiv \left( {3/2}\right) ^{1/3}\left( 9\gamma ^{2}|\Phi |^{4}+\sqrt{3}%
\sqrt{256+27\gamma ^{4}|\Phi |^{8}}\right) ^{1/3}.  \label{A}
\end{equation}%
The dependence of $\eta $ on product $\gamma |\Phi |^{2},$ as produced by
Eqs. (\ref{eta-solve}) and (\ref{A}), is displayed in Fig. \ref{fig_eta}.

%fig1

\subsection{Low- and high-density limits}

In the low-density limit, $|\gamma ||\Phi |^{2}\ll 1$, Eq. (\ref{eta})
yields a simple solution,%
\begin{equation}
\eta _{\mathrm{low}}\approx 1+\gamma |\Phi |^{2}/4\text{.}  \label{low}
\end{equation}%
The substitution of this approximation in Eq. (\ref{npse}) reduces it to the
nonlinear Schr\"{o}dinger equation (NPSE) with the cubic-quintic (CQ)
nonlinearity, in which the quintic term always corresponds to the effective
self-attraction:%
\begin{equation}
i{\frac{\partial }{\partial t}}\Phi =\left[ -{\frac{1}{2}}\nabla _{\bot
}^{2}+W(x,y)+\gamma |\Phi |^{2}-{\frac{3}{16}}\gamma ^{2}|\Phi |^{4}\right]
\Phi .  \label{CQ}
\end{equation}%
The derivation of the effective 1D equation from the 3D GPE in the
low-density limit also leads to the NPSE with the CQ nonlinearity of the
same type \cite{1D}. Note that the self-focusing quintic term in the 2D NPSE
leads to the supercritical collapse (the one with zero threshold). If the 2D
equation includes both quintic and cubic self-attraction terms, then a
periodic potential, accounted for by $W(x,y)$ in Eq. (\ref{CQ}), can
stabilize various types of solitons and localized vortices against the
(supercritical) collapse \cite{Radik}. However, the 2D equation with the
quintic nonlinearity only, see Eq. (\ref{quint}) below, cannot give rise to
stable solitons, irrespective of the presence of the periodic potential \cite%
{Radik}.

In the high-density limit, $|\gamma ||\Phi |^{2}\gg 1$, asymptotic
expressions for the axial width which follow from Eq. (\ref{eta}) are
different in the cases of repulsion ($\gamma >0$) and attraction ($\gamma <0$%
),
\begin{equation}
\eta _{\mathrm{high}}^{\mathrm{(rep)}}\approx (\gamma |\phi
|^{2})^{1/3},\,\eta _{\mathrm{high}}^{\mathrm{(attr)}}=-\left( \gamma
\left\vert \Phi \right\vert ^{2}\right) ^{-1}.  \label{high}
\end{equation}%
The substitution of these approximations in Eq. (\ref{npse}) leads to two
different 2D NPSEs: in the case of the repulsion, it is%
\begin{equation}
i{\frac{\partial }{\partial t}}\Phi =\left[ -{\frac{1}{2}}\nabla _{\bot
}^{2}+W(x,y)+{\frac{5}{4}}\gamma ^{2/3}|\Phi |^{4/3}\right] \Phi ,
\label{4/3}
\end{equation}%
which coincides with the equation of the hydrodynamic type for the
degenerate fermion gas in the weak-coupling regime \cite{SKA}, and in the
case of the attraction, the effective equation is the NPSE with the quintic
term only:
\begin{equation}
i{\frac{\partial }{\partial t}}\Phi =\left[ -{\frac{1}{2}}\left( {\frac{%
\partial ^{2}}{\partial x^{2}}}+{\frac{\partial ^{2}}{\partial y^{2}}}%
\right) +W(x,y)-\frac{3}{4}\gamma ^{2}\left\vert \Phi \right\vert ^{4}\right]
\Phi .  \label{quint}
\end{equation}

\subsection{The Thomas-Fermi approximation}

The substitution of $\Phi (x,y,t)=\phi (x,y)\ e^{-i\bar{\mu}t},$ with
chemical potential $\bar{\mu}$, casts Eq. (\ref{npse}) in the stationary
form,
\begin{equation}
\left[ -{\frac{1}{2}}\nabla _{\bot }^{2}+{\frac{\gamma |\phi |^{2}}{\eta }}+{%
\frac{1}{4}}\left( {\frac{1}{\eta ^{2}}}+\eta ^{2}\right) \right] \phi =\mu
(x,y)\phi ,  \label{stat}
\end{equation}%
where $\mu (x,y)\equiv \bar{\mu}-W(x,y)\;$is the local chemical potential.
In the Thomas-Fermi (TF) approximation, that may be relevant in the case of
the repulsive interactions, i.e., $\gamma >0$, one neglects the spatial
derivatives in Eq. (\ref{stat}), and, taking into regard Eq. (\ref{eta-solve}%
), obtains the following expression,
\begin{equation}
|\phi (x,y)|^{2}={\frac{{\frac{4}{25}}\left\{ \sqrt{\left[ \mu (x,y)\right]
^{2}+{\frac{15}{4}}}+\mu (x,y)\right\} ^{2}-1}{\gamma \sqrt{\frac{2}{5}%
\left\{ \sqrt{\left[ \mu (x,y)\right] ^{2}+{\frac{15}{4}}}+\mu (x,y)\right\}
}}},  \label{FP}
\end{equation}%
which is valid as long as it yields $|\phi (x,y)|^{2}>0$. In the region
where expression (\ref{FP}) is negative, it is replaced by $|\phi
(x,y)|^{2}\equiv 0$. In the case of $\left\vert \mu \left( x,y\right)
-1/2\right\vert \ll 1/2$, the TF approximation corresponds to the
low-density limit (\ref{low}), with $|\phi (x,y)|^{2}\approx \gamma ^{-1}%
\left[ \mu (x,y)-{1/2}\right] .$ The high-density limit (\ref{high})
corresponds to large $\mu $, which yields $|\phi (x,y)|^{2}\approx {\gamma }%
^{-1}\left[ \left( 4/5\right) \mu (x,y)\right] ^{3/2}.$

\section{The ground state and vortices in the 2D parabolic trap, with the
repulsive nonlinearity}

We start the analysis of trapped states, predicted by the 2D NPSE, with the
case of the repulsion, assuming that the potential inside the ``pancake" is
also a parabolic trap (which is much weaker than the underlying axial
trapping potential, $z^{2}/2$), i.e.,
\begin{equation}
W(x,y)={(\lambda }^{2}{/2)}(x^{2}+y^{2})  \label{lambda}
\end{equation}%
in Eqs. (\ref{npse}) and (\ref{stat}), with small $\lambda $. For comparison
of localized states predicted by stationary NPSE (\ref{stat}), and by the
full 3D\ GPE, we introduce the radial and axial probability densities,
defined in terms of the 3D wave function, which is subject to normalization (%
\ref{norm}):%
\begin{equation}
\rho _{\mathrm{3D}}(r,t)=\int_{-\infty }^{+\infty }|\Psi (\mathbf{r}%
,t)|^{2}\ dz,~\rho _{\mathrm{3D}}(z,t)=2\pi \int_{0}^{\infty }|\Psi (\mathbf{%
r},t)|^{2}rdr,  \label{densities}
\end{equation}%
where $r\equiv \sqrt{x^{2}+y^{2}}$. In the framework of the 2D description,
the factorized ansatz based on Eqs. (\ref{factor}) and (\ref{Xi}) yields%
\begin{equation}
\rho _{\mathrm{2D}}(r,t)=|\Phi (r,t)|^{2},~\rho _{\mathrm{2D}}(z,t)=2\sqrt{%
\pi }\int_{0}^{\infty }\frac{rdr}{\eta (r)}e^{-z^{2}/\eta ^{2}(r)}\Phi
^{2}(r).  \label{2Ddensity}
\end{equation}

\begin{figure}[tbp]
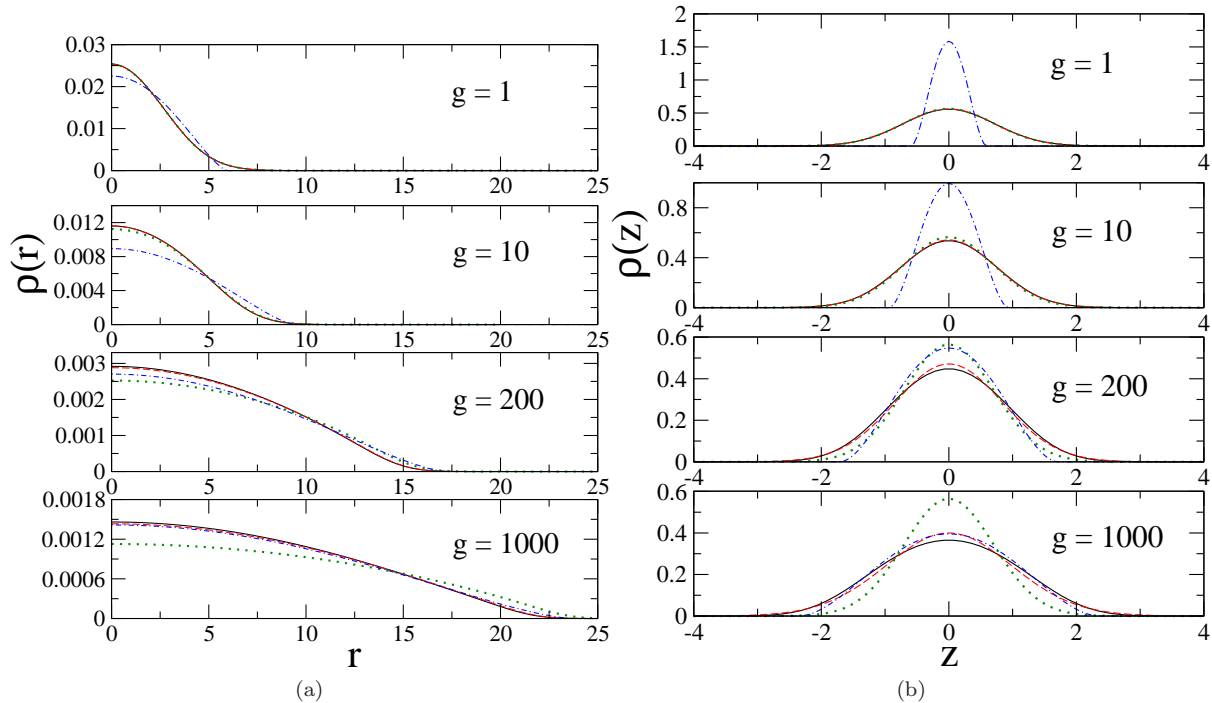

\subfigure[]{\includegraphics[width=8.cm,clip]{2dnpse-f2a.eps}}{%
\subfigure[]{\includegraphics[width=8.cm,clip]{2dnpse-f2b.eps}}}
\caption{(Color online). Radial (a) and axial (b) probability densities, $%
\protect\rho (r)$ and $\protect\rho (z)$, for the self-repulsive condensate
in the ground state trapped in parabolic potential (\protect\ref{lambda})
with $\protect\lambda =0.1$. Solid, dashed, dotted, and dotted-dashed curves
display the results produced, respectively, by the 3D GPE, 2D NPSE, the
ordinary 2D equation with the cubic nonlinearity, and TF approximation to
the 3D GPE. Values of interaction strength $g=2Na_{s}/a_{z}$ are indicated
in the panels.}
\label{fig2}
\end{figure}

In Fig. \ref{fig2}, we display typical examples of the radial and axial
densities for the repulsive condensate in the ground state, fixing $\lambda
=0.1$. The respective 3D and 2D stationary wave functions were obtained by
means of the imaginary-time integration, using the finite-difference
Crank-Nicolson predictor-corrector code, as elaborated in Ref. \cite%
{sala-numerics}. The figure demonstrates that the 2D NPSE provides almost
exact results, in the comparison to the full 3D equation, except for some
deviation of the axial density at very large values of the interaction
strength, $g=200$ and $1000$. It is worthy to note that the NPSE provides
for a much better accuracy than the formal 2D GPE with the cubic
nonlinearity.

%fig2

In the same setting, the 3D wave functions for vortex states can be sought
for as $\Psi (\mathbf{r},t)=\psi (r,z,t)\ e^{iS\theta },$where $r$ and $%
\theta $ are the polar coordinates in the $\left( x,y\right) $ plane, $S$ is
integer vorticity, and function $\psi $ obeys equation
\begin{eqnarray}
i{\frac{\partial }{\partial t}}\psi  &=&\frac{1}{2}\left[ -\left( {\frac{%
\partial ^{2}}{\partial r^{2}}}+{\frac{1}{r}}{\frac{\partial }{\partial r}}+{%
\frac{\partial ^{2}}{\partial z^{2}}}\right) +{\frac{S^{2}}{r^{2}}}\right.
\notag \\
&&\left. +\lambda ^{2}r^{2}+z^{2}+4\pi g|\psi |^{2}\right] \psi \;.
\label{3dgpe-vortex}
\end{eqnarray}%
In the case of 2D NPSE (\ref{npse}), the vortex corresponds to $\Phi
(r,\theta ,t)=\phi (r,t)e^{iS\theta },$ with Eq. (\ref{3dgpe-vortex})
replaced by
\begin{gather}
i{\frac{\partial }{\partial t}}\phi =\frac{1}{2}\left[ -\left( {\frac{%
\partial ^{2}}{\partial r^{2}}}+{\frac{1}{r}}{\frac{\partial }{\partial r}}%
\right) +{\frac{S^{2}}{r^{2}}}\right.   \notag \\
\left. +\lambda ^{2}r^{2}+2\gamma {\frac{|\phi |^{2}}{\eta }}+{\frac{1}{2}}%
\left( {\frac{1}{\eta ^{2}}}+\eta ^{2}\right) \right] \phi .
\label{npse-vortex}
\end{gather}%
Relevant solutions to both equations (%
\index{3dgpe-vortex}) and (\ref{npse-vortex}) must feature the usual
asymptotic form, $\left( \psi ,\phi \right) \sim r^{|S|}$, at $r\rightarrow 0
$.

\begin{figure}[b]
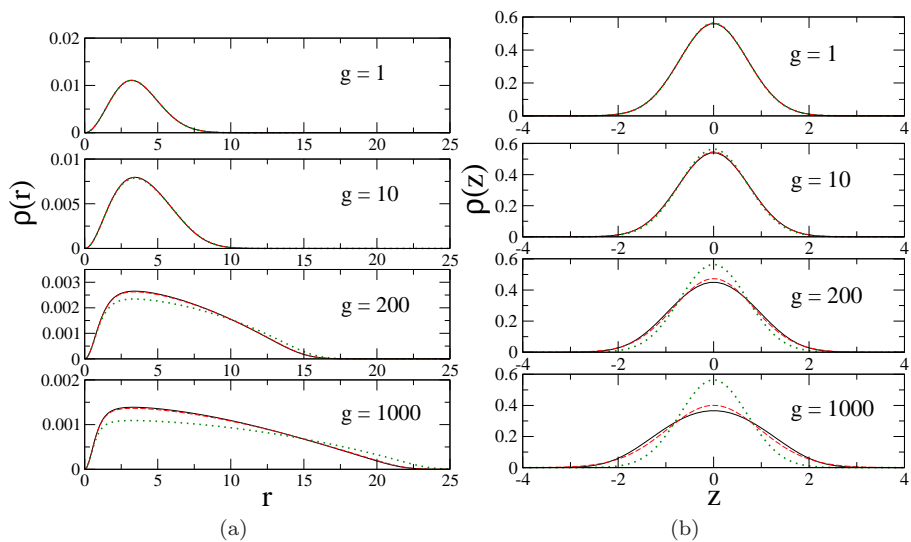

{\subfigure[]{\includegraphics[width=6.cm,clip]{2dnpse-f3a.eps}}}%
\subfigure[]{\includegraphics[width=6.cm,clip]{2dnpse-f3b.eps}}
\caption{(Color online). The same as in Fig. \protect\ref{fig2}, but for
vortex states with $S=1$.}
\label{fig3}
\end{figure}

The radial and axial profiles of the vortex states with $S=1$ are displayed
in Fig. \ref{fig3}, for the same value of $\lambda =0.1$ as in Fig. \ref%
{fig2}. To generate these results, Eqs. (\ref{3dgpe-vortex}) and (\ref%
{npse-vortex}) were also solved in the imaginary time by means of the
above-mentioned finite-difference Crank-Nicolson algorithm \cite%
{sala-numerics}. We again conclude that, as well as in the case of the
ground state, the 2D NPSE produces results which, in most cases, virtually
coincide with those obtained from the 3D GPE (except for some deviation of
the axial profiles at very large $g$), while the formal 2D GPE with the
cubic nonlinear term is essentially less accurate.

%fig3

\begin{figure}[tbp]
{\includegraphics[width=6.cm,clip]{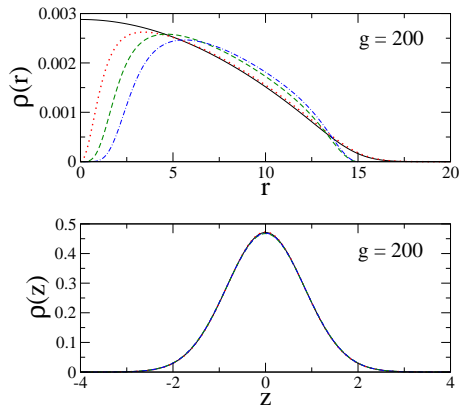}}
\caption{(Color online). The upper and lower panels display, respectively,
the radial and axial probability densities, $\protect\rho (r)$ and $\protect%
\rho (z)$, for the ground state and higher-order vortices, as found from the
2D NPSE with $\protect\lambda =0.1$ and $g\equiv 2Na_{s}/a_{z}=200$. The
solid, dotted, dashed, and dotted-dashed curves pertain, respectively, to $%
S=0$, $S=1$, $S=2$, and $S=3$ (all vortices with $S\geq 2$ are unstable
against splitting into unitary vortices).}
\label{fig4}
\end{figure}

For higher values of $S$ (up to $S=3$), the radial and axial density
profiles, as predicted by the 2D NPSE, are displayed, along with their
counterparts for the ground state, in Fig. \ref{fig4}, at fixed values of
the parameters, \ $\lambda =0.1$ and $g=200$. As expected, the growth of $S$
leads to the increase of the size of the vortex' core. On the other hand,
the axial density profile, $\rho (z)$, is not strongly affected by the
change of $S$.

We have checked that, as well as in the case of full 3D GPE with the
repulsive nonlinearity and its 2D counterpart, multiple trapped vortices are
unstable against splitting into solitary vortices within the framework of
the 2D NPSE. This finding is not surprising, as the latter equation is
derived from the 3D GPE, hence it should inherit basic properties of
solutions of the 3D equation.

%fig4

\section{Solitons and solitary vortices in the case of the attractive
nonlinearity}

\subsection{The axisymmetric parabolic trap}

\begin{figure}[tbp]
{\includegraphics[width=8.cm,clip]{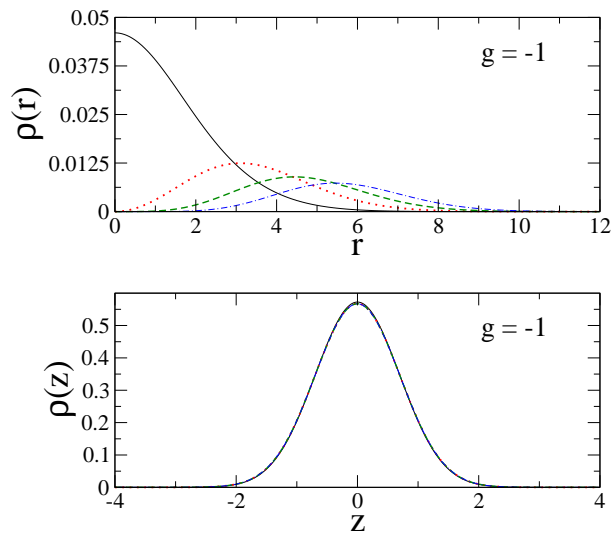}}
\caption{(Color online). The same as in Fig. \protect\ref{fig4} (again, with
$\protect\lambda =0.1$), but for the self-attractive condensate, with
nonlinearity strength $g\equiv 2Na_{s}/a_{z}=-1$.}
\label{fig5}
\end{figure}

In the case of the attractive nonlinearity, $g<0$, localized states and
vortices supported by the in-plane potential -- in particular, axisymmetric
parabolic potential (\ref{lambda}) -- may be considered as solitons \cite%
{Dodd}-\cite{we-vortex-in-tube}. In Fig. \ref{fig5} we display the radial
and axial profiles of the solitons found from the NPSE with potential (\ref%
{lambda}), for $\lambda =0.1$ and $g=-1$. It can be checked that, for
sufficiently small values of $|g|$, at which the collapse does \ not take
place, the full 3D GPE yields virtually the same profiles.

%fig5

Under normalization conditions (\ref{norm}) and parabolic confinement (\ref%
{lambda}), the critical value, $g_{c}$, of the interaction strength for the
onset of collapse can be found as that at which the stationary soliton
solution disappears. These values, as found from the set of the three models
(3D GPE, 2D NPSE, and the formal 2D GPE, with $\lambda =0.1$), are collected
in Table 1, for different values of vorticity $S$. Naturally, $g_{c}$
quickly increases with $S$ \cite{DumDum}, and we once again conclude that
the NPSE provides for high accuracy, especially in comparison with the large
error in the predictions produced by the formal 2D equation with the
attractive cubic nonlinearity.

\bigskip

\begin{center}
\begin{tabular}{|c|c|c|c|}
\hline
~~~$S$~~~ & ~~$\left( ~g_{c}\right) _{\mathrm{3D-GPE}}$~~~ & ~~~$\left(
g_{c}\right) _{\mathrm{2D-NPSE}}$~~~ & ~~~$\left( g_{c}\right) _{\mathrm{%
2D-GPE}}$~~ \\ \hline
~~~$0$~~~ & ~~~$-1.9$~~~ & ~~~$-1.9$~~~ & ~~~$-2.3$~~~ \\
~~~$1$~~~ & ~~~$-7.3$~~~ & ~~~$-7.2$~~~ & ~~~$-8.9$~~~ \\
~~~$2$~~~ & ~~~$-12.4$~~~ & ~~~$-12.3$~~~ & ~~~$-15.8$~~~ \\
~~~$3$~~~ & ~~~$-16.9$~~~ & ~~~$-16.8$~~~ & ~~~$-21.6$~~~ \\ \hline
\end{tabular}
\end{center}

\noindent Table 1. Critical strength $g_{c}$ for the onset of the collapse
in the attractive condensate with vorticity $S$, confined by in-plane
potential (\ref{lambda}) with $\lambda =0.1$. The table compares the
predictions produced by the full 3D GPE, the effective 2D equation (NPSE),
and the formal 2D equation (GPE) with the cubic nonlinearity. \bigskip

An important issue is the stability of solitary vortices with vorticity $S$
in the axisymmetric harmonic trap. We tackled this problem by means of
direct simulations of the time-dependent 2D NPSE, Eq. (\ref{npse}), using a
finite-difference real-time Crank-Nicolson algorithm \cite{sala-numerics}.
The initial conditions were taken as%
\begin{equation}
\Phi (x,y,t=0)=C\ (x+iy)^{S}\ \exp \left[ -%
\frac{\lambda }{2}\left( x^{2}+\frac{y^{2}}{\delta ^{2}}\right) \right] ,
\label{pippo}
\end{equation}%
where $C$ is a normalization constant, and $\delta \neq 1$ accounts for
possible breakup of the axial symmetry of the vortex, i.e., an initial
perturbation applied to the vortex. Note that, with $\delta =1$, Eq. (\ref%
{pippo}) gives the exact quantum-mechanical wave function of the stationary
vortex configuration (for $g=0$).

We report results of the simulations for $\delta =1.1$ and trap's strength $%
\lambda =0.1$. We have concluded that the trapped vortex with $S=1$ is
stable for $g>-1.3$: in this region, the initial perturbation leads only to
intrinsic periodic oscillations of the vortex, which maintains its
integrity. In agreement with previously published results \cite{Ueda}-\cite%
{DumDum2}, the value of the nonlinearity strength at the stability border, $%
g_{\mathrm{stab}}=-1.3$, is much lower than at the collapse threshold, $%
g_{c}=-7.2$.

\begin{figure}[tbp]
{\includegraphics[width=6.cm,angle=-90,clip]{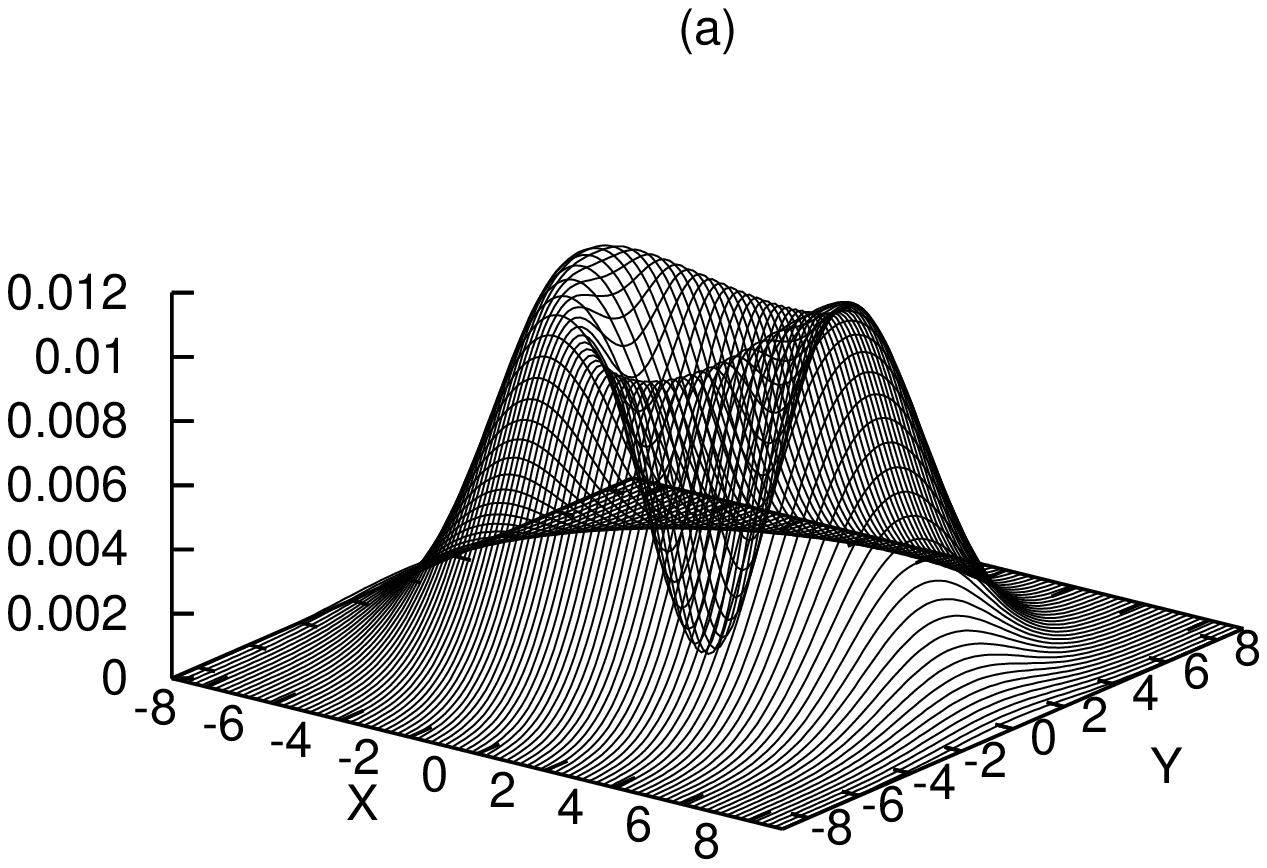} %
\includegraphics[width=6.cm,angle=-90,clip]{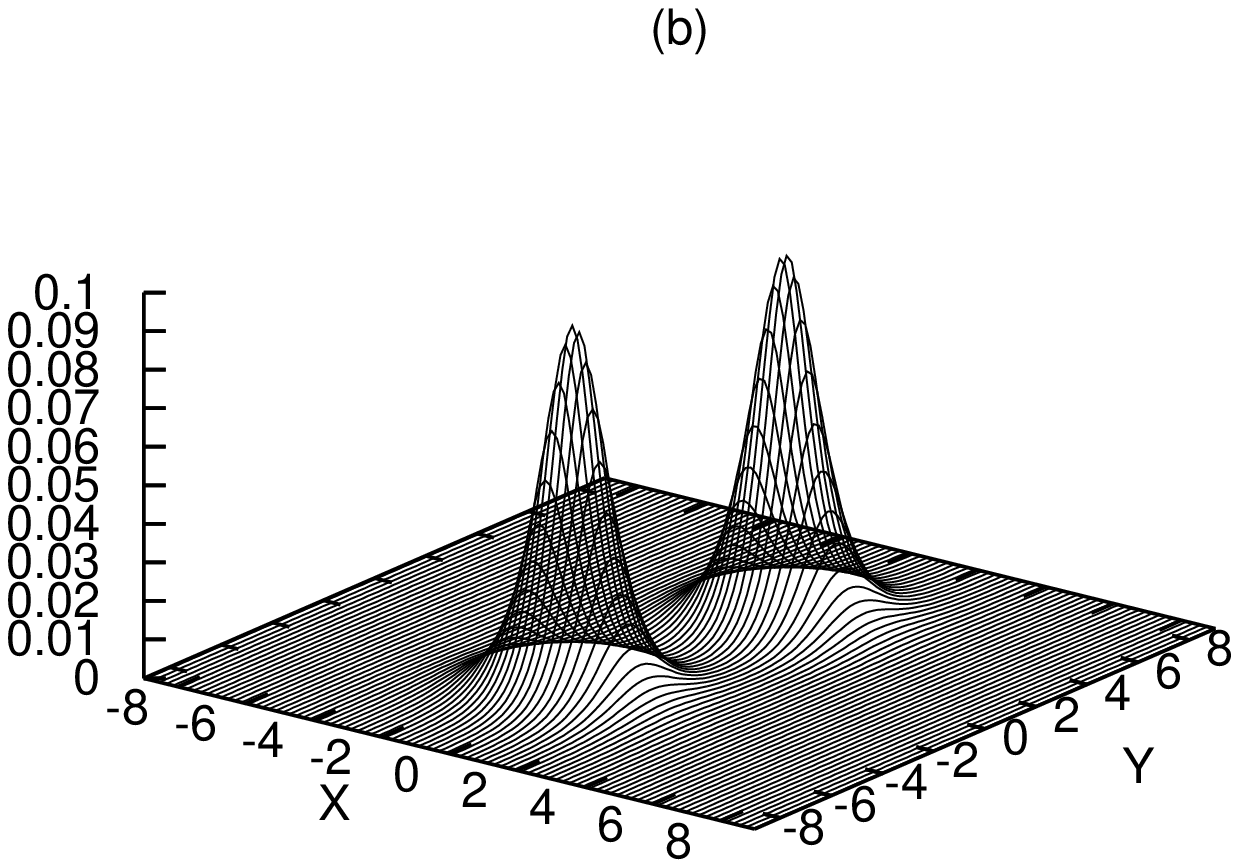} %
\includegraphics[width=6.cm,angle=-90,clip]{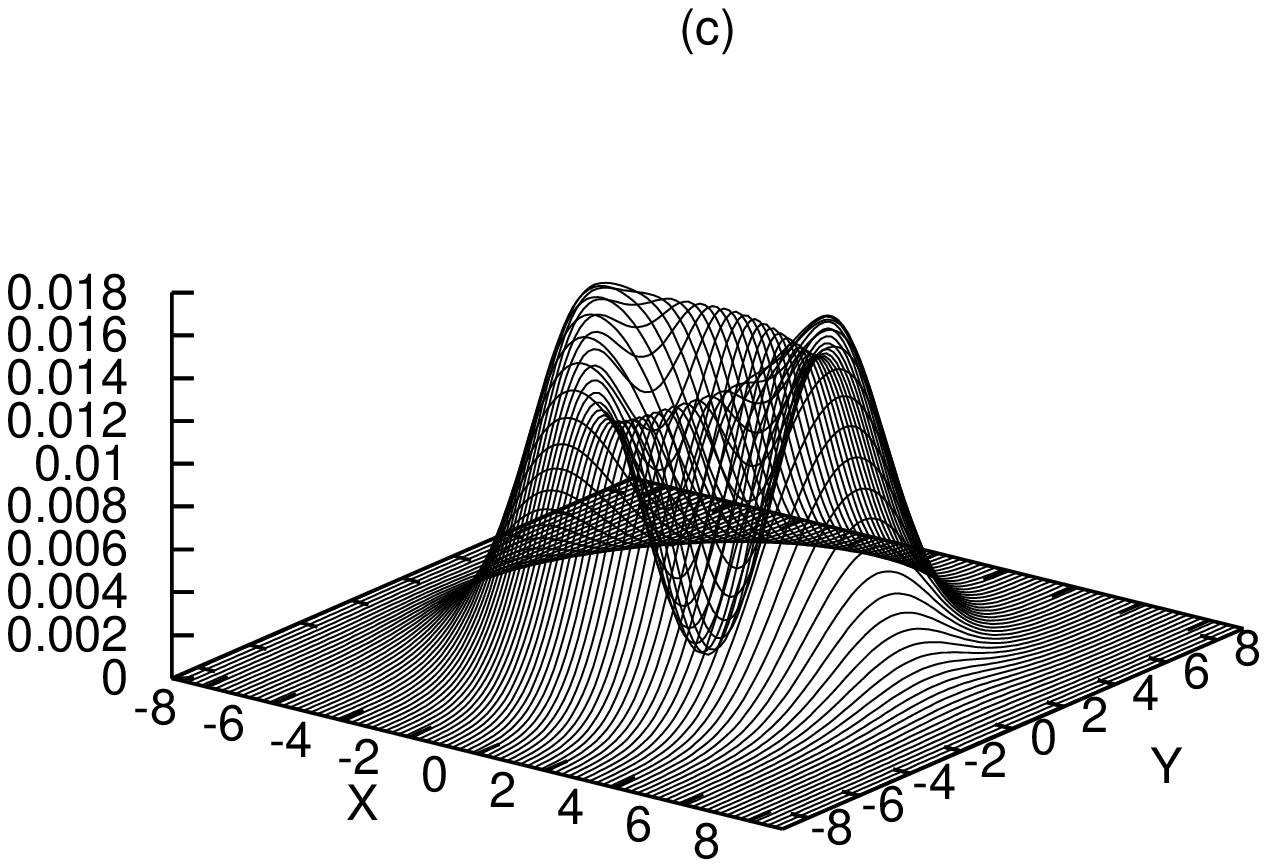} %
\includegraphics[width=6.cm,angle=-90,clip]{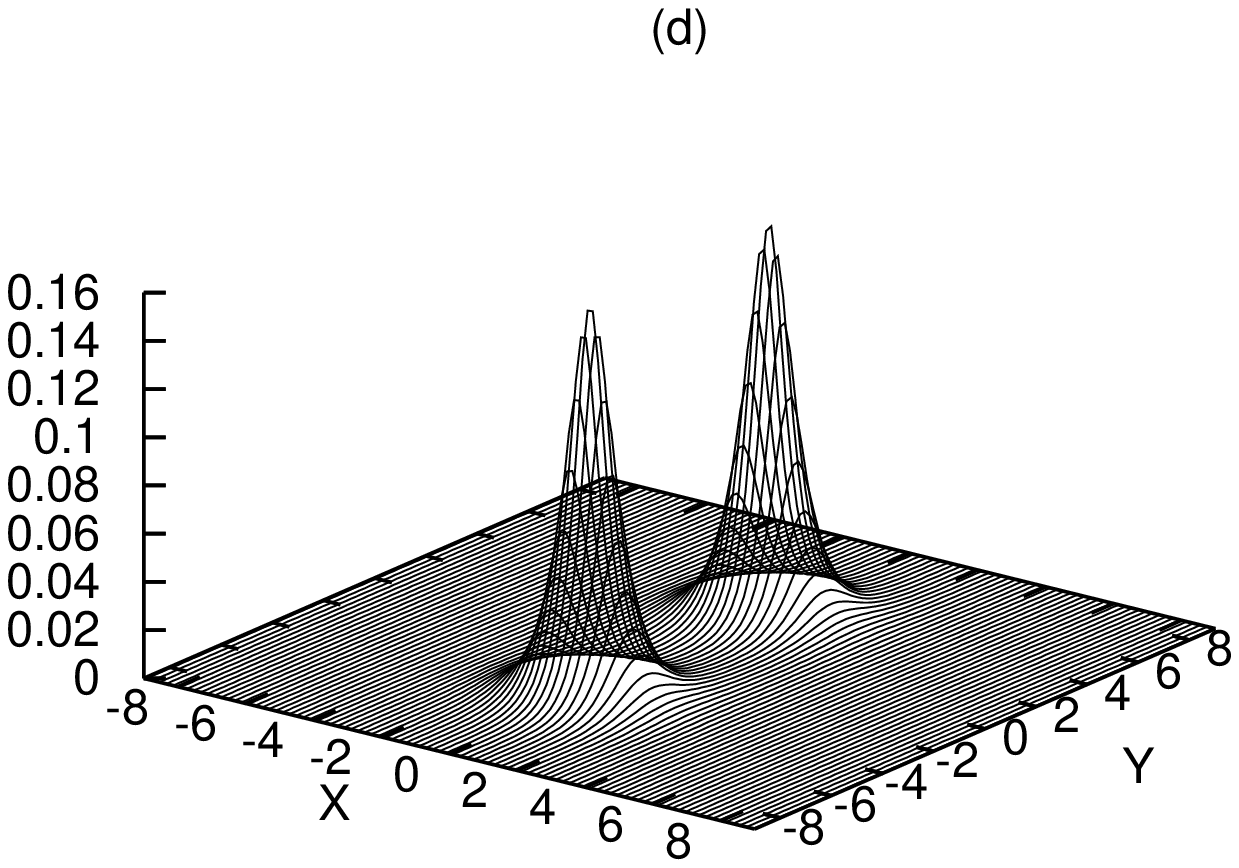}}
\caption{Density profiles $|\Phi (x,y,t)|^{2}$ for the solitary vortex with $%
S=1$ and $g=-1.4$, obtained by solving the time-dependent 2D NPSE, Eq. (%
\protect\ref{npse}), with $\protect\lambda =0.1$. The four snapshots
correspond to $t=0$ (a), $t=56$ (b), $t=88$ (c), $t=120$ (d). Time is given
in units of $1/\protect\omega _{z}$. The figure displays the quasiperiodic
process of the alternating splittings and recombinations of the vortex
ending with the collapse, see the text.}
\label{fig6}
\end{figure}

Outside the stability domain, i.e., at $-g\geq 1.3$, we observed splitting
of the trapped vortex into two solitary fragments, which would rotate and
recombine back into the vortex. In the interval $1.3\leq -g< 1.4$ the
splitting-recombination cycle repeats itself quasi-periodically, as shown in
Fig. \ref{fig6}, where we have plotted the planar density profile, $|\Phi
(x,y,t)|^{2}$, at four instants of time corresponding to different stages of
the cycle. Note that a similar dynamical regime was previously demonstrated
in Refs. \cite{Ueda,DumDum}, in the framework of the ordinary 2D equation
with the cubic self-attractive term. If the self-attraction is stronger,
\textit{viz}., $1.4\leq -g < 1.5$, the two fragments suffer an intrinsic
collapse after two cycles (in the case shown in Fig. \ref{fig6} the collapse
takes place at $t\simeq 125$). For a still stronger self-attraction, $-g\geq
1.5$, the two rotating fragments do not recombine at all, quickly undergoing
the intrinsic collapse.

%fig6

\begin{figure}[tbp]
{\includegraphics[width=6.cm,angle=-90,clip]{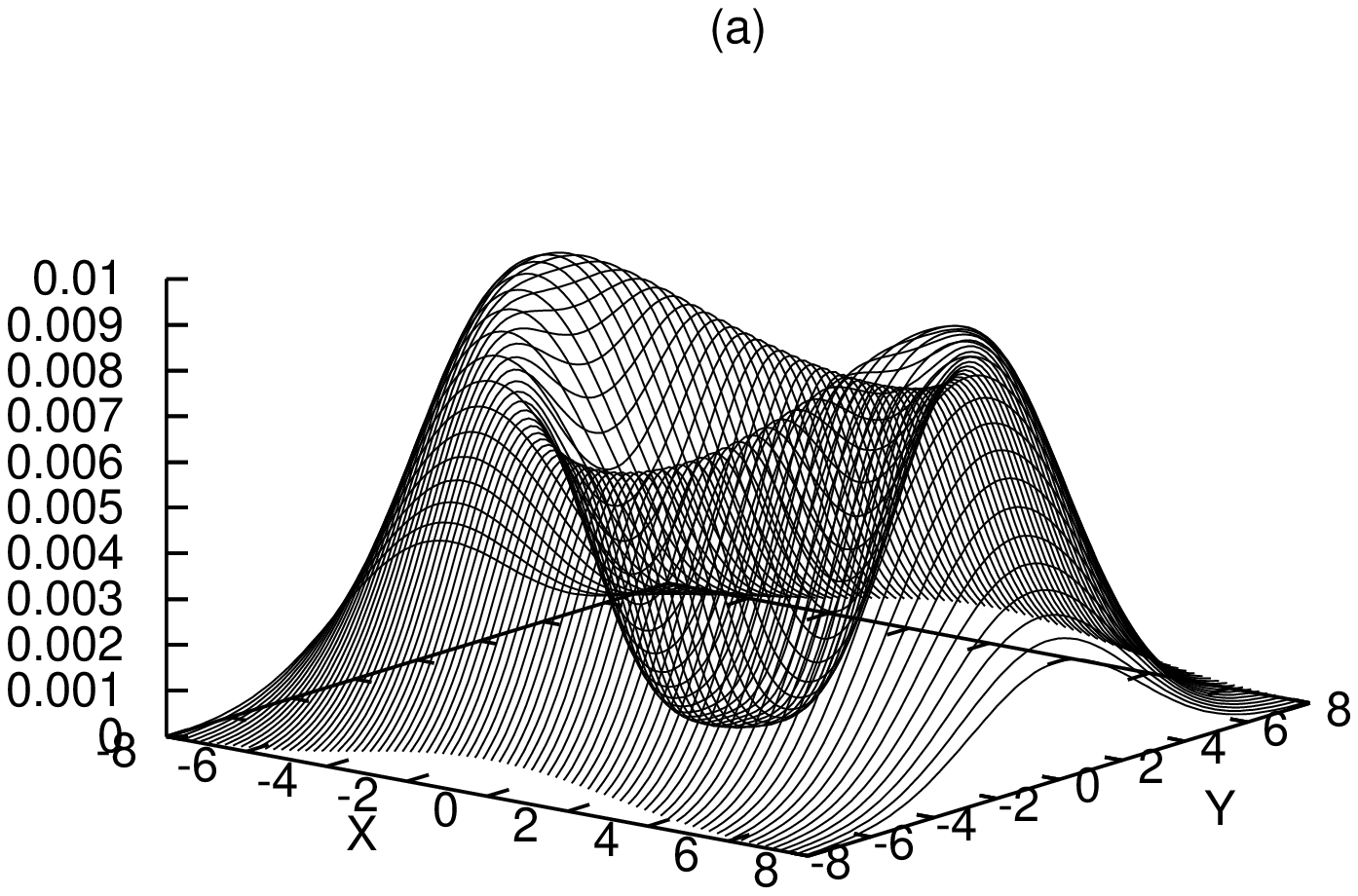} %
\includegraphics[width=6.cm,angle=-90,clip]{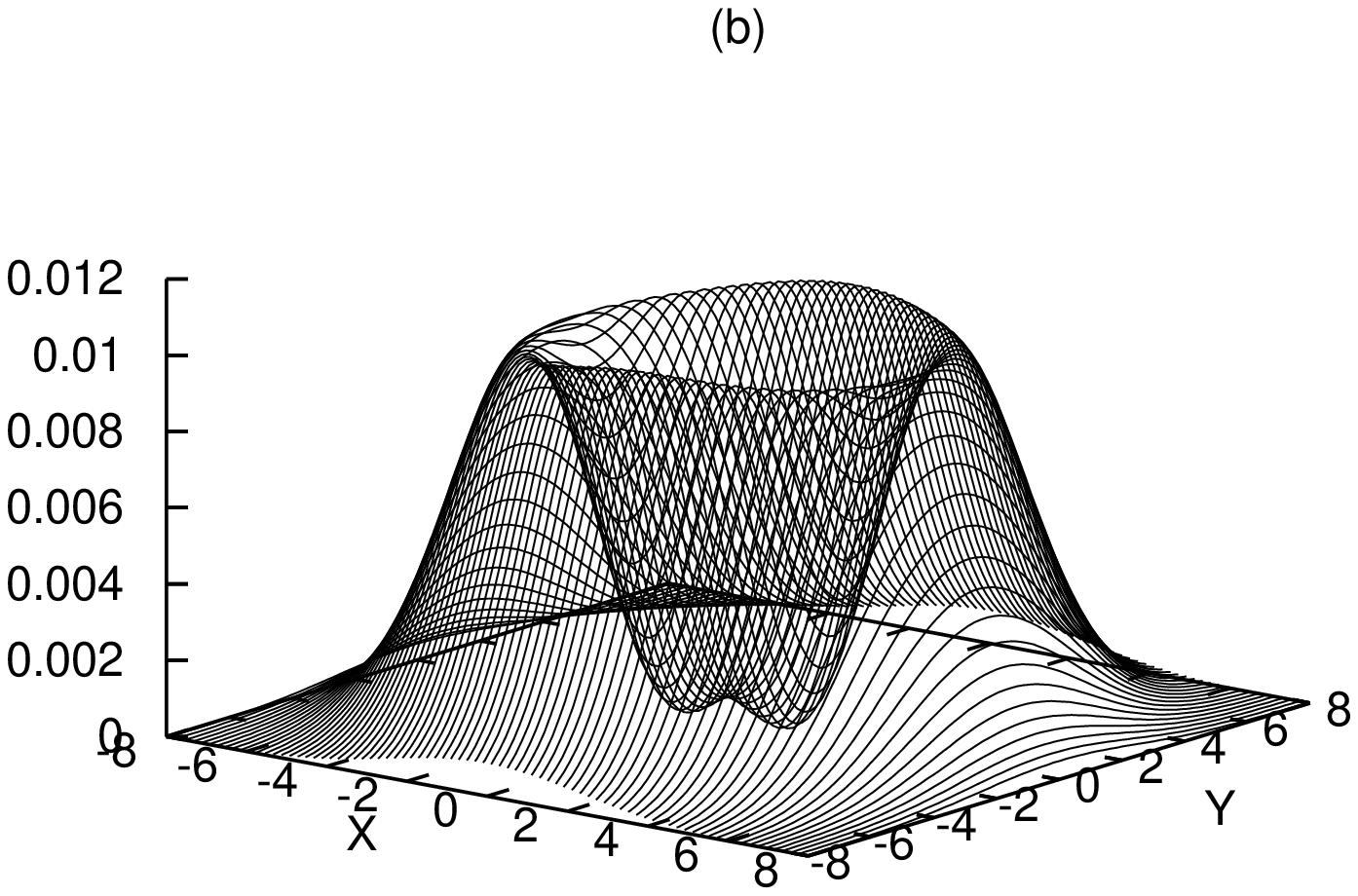} %
\includegraphics[width=6.cm,angle=-90,clip]{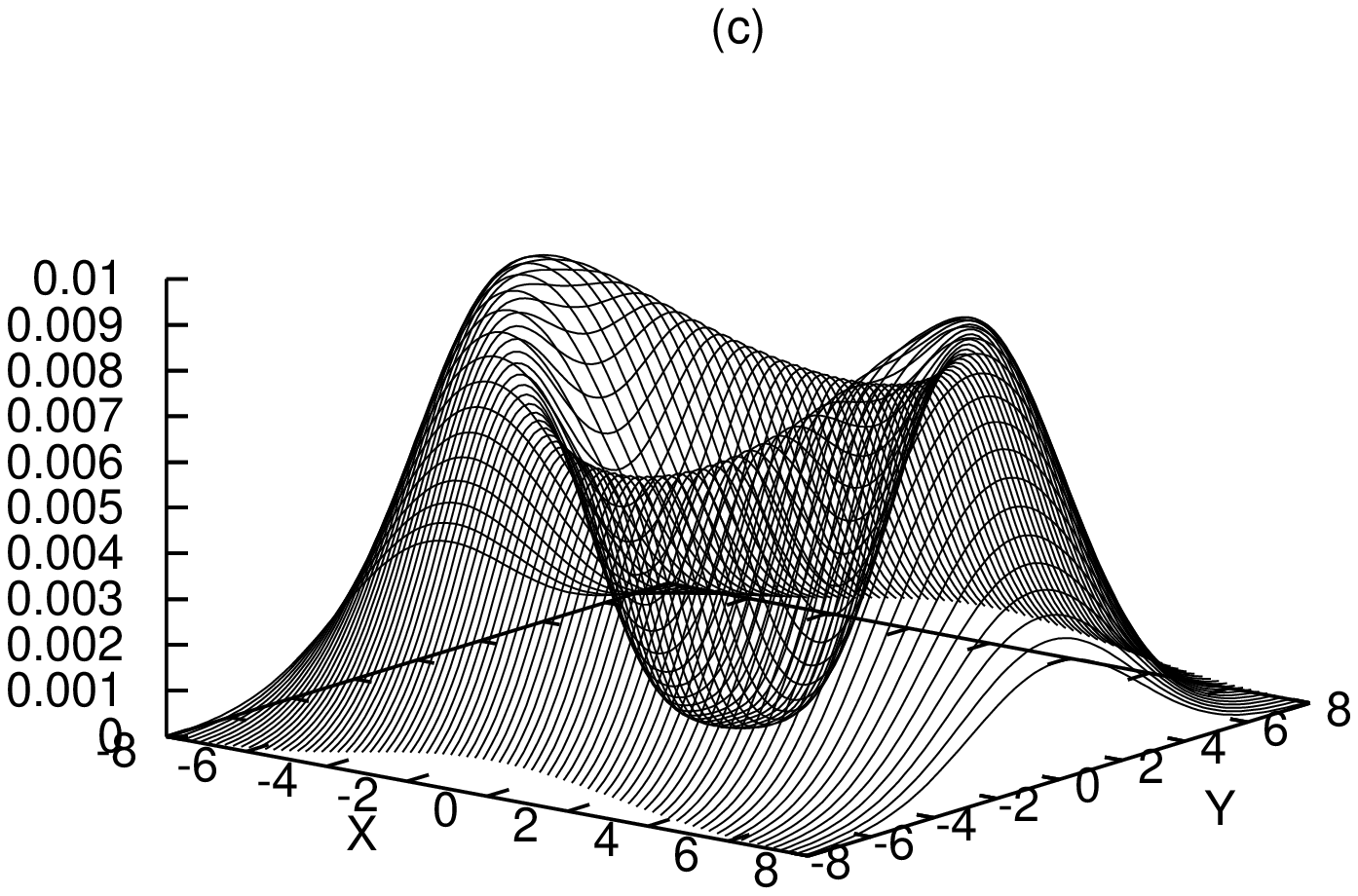} %
\includegraphics[width=6.cm,angle=-90,clip]{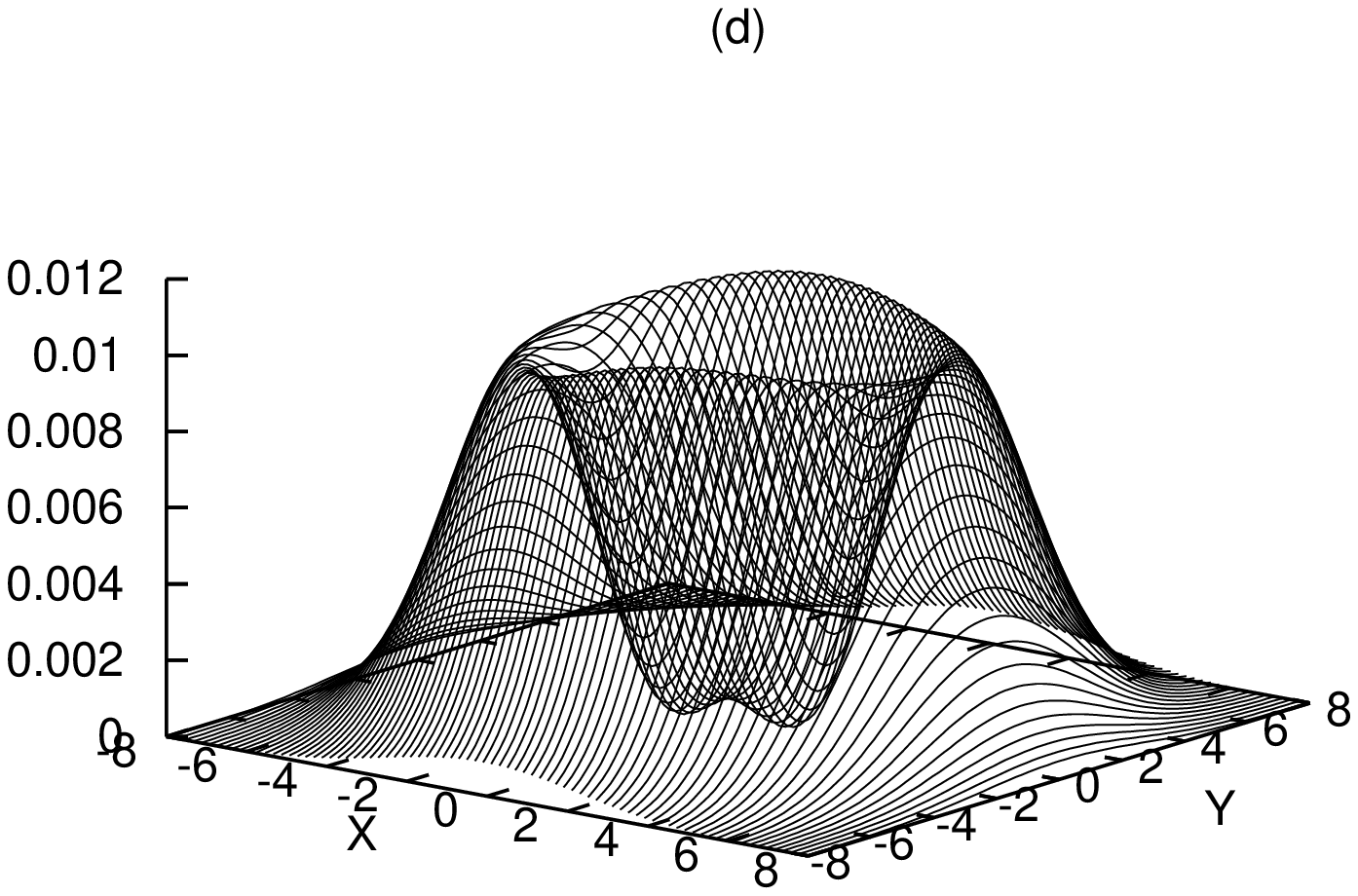}}
\caption{The same as in Fig. \protect\ref{fig6}, but for the solitary vortex
with $S=2$ and $g=-0.3$. The snapshots pertain to $t=0$ (a), $t=16$ (b), $%
t=32$ (c), $t=48$ (d). The figure displays one-and-a-half cycles of the
periodic process (of period $\protect\tau\simeq 32$) of the alternating
splittings and recombinations of the vortex, see the text.}
\label{fig7}
\end{figure}

We have also investigated the stability of the solitary vortex with $S=2$.
In agreement with Ref. \cite{DumDum}, it was found that the double vortex
can never be stable against splitting into a pair of unitary ones.
Nevertheless, a stable dynamical regime, which has not been reported in
previous studies, was found in this case: the simulations of Eq. (\ref{npse}%
), with initial conditions (\ref{pippo}) corresponding to $S=2$, demonstrate
that, at small values of the interaction strength, $-g~\leq 0.5$, the double
vortex periodically features incomplete separation into two vortices, which
manifests itself in the splitting of the zero-amplitude point (the center of
the vortex) into a pair of local minima, separated by a relatively small
distance [see Figs. \ref{fig7}(b,d)], which is followed by recombination of
the pair back into a single point, see a set of snapshots in Fig. \ref{fig7}.

%fig7

\begin{figure}[tbp]
{\includegraphics[width=6.cm,angle=-90,clip]{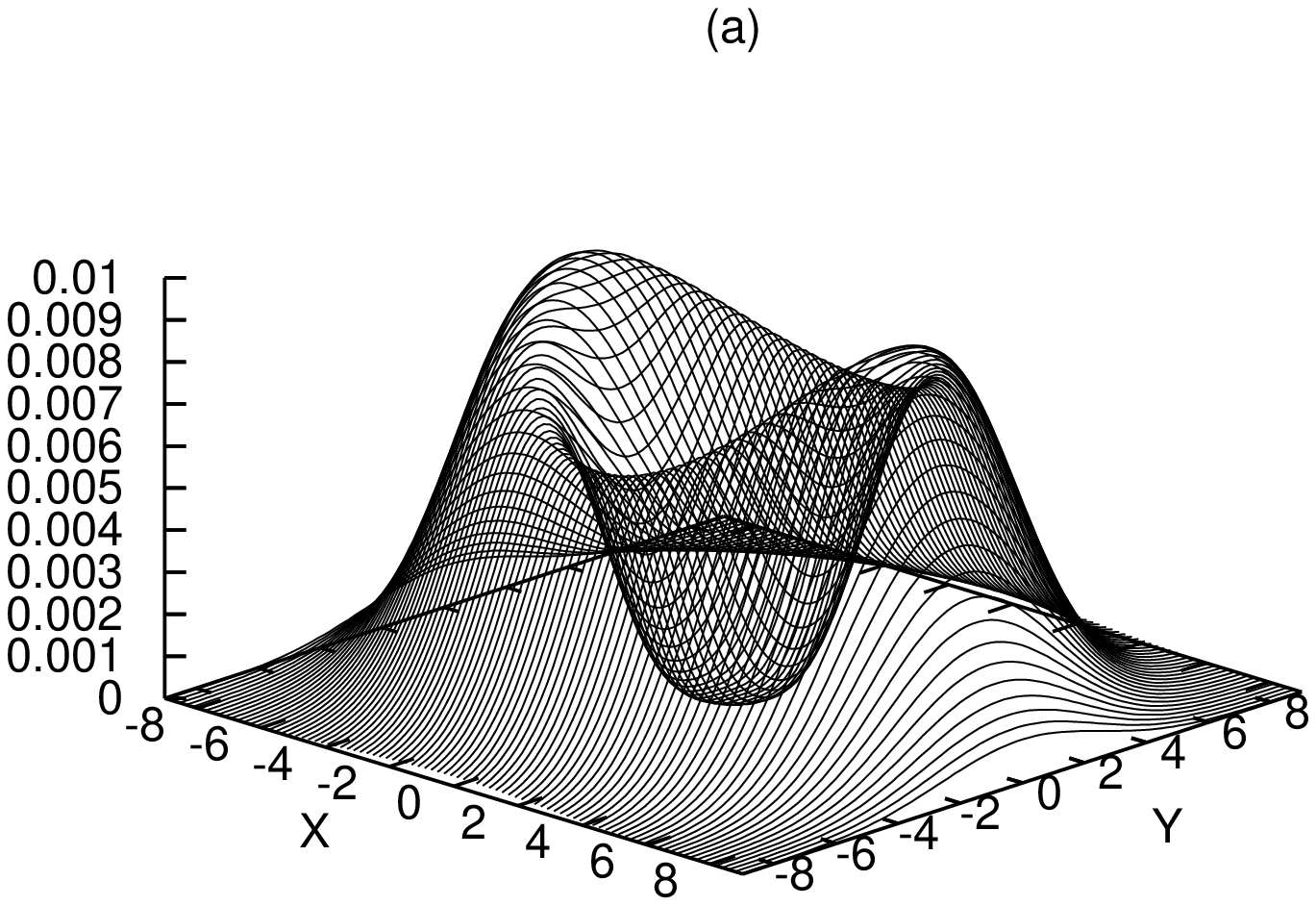} %
\includegraphics[width=6.cm,angle=-90,clip]{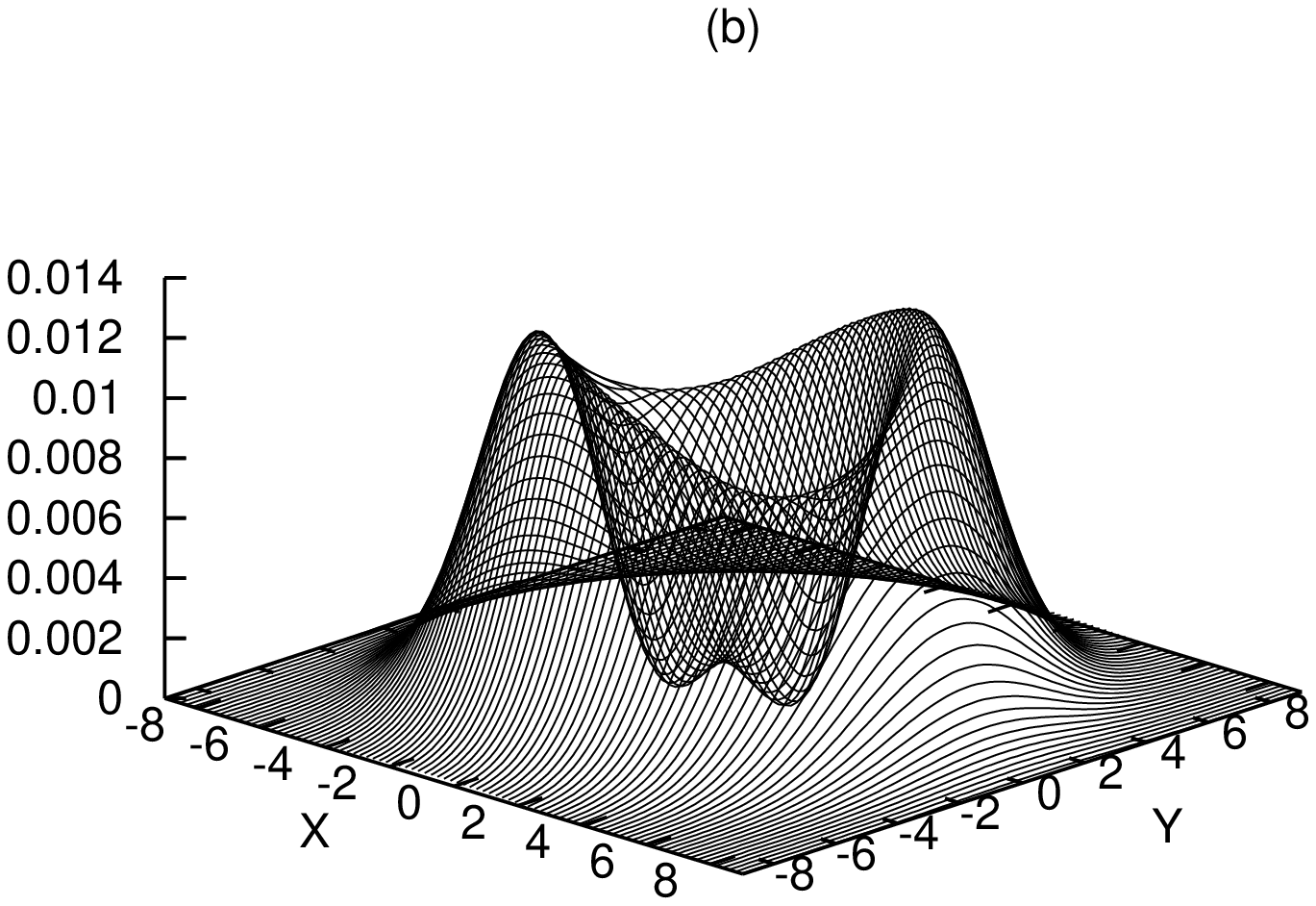} %
\includegraphics[width=6.cm,angle=-90,clip]{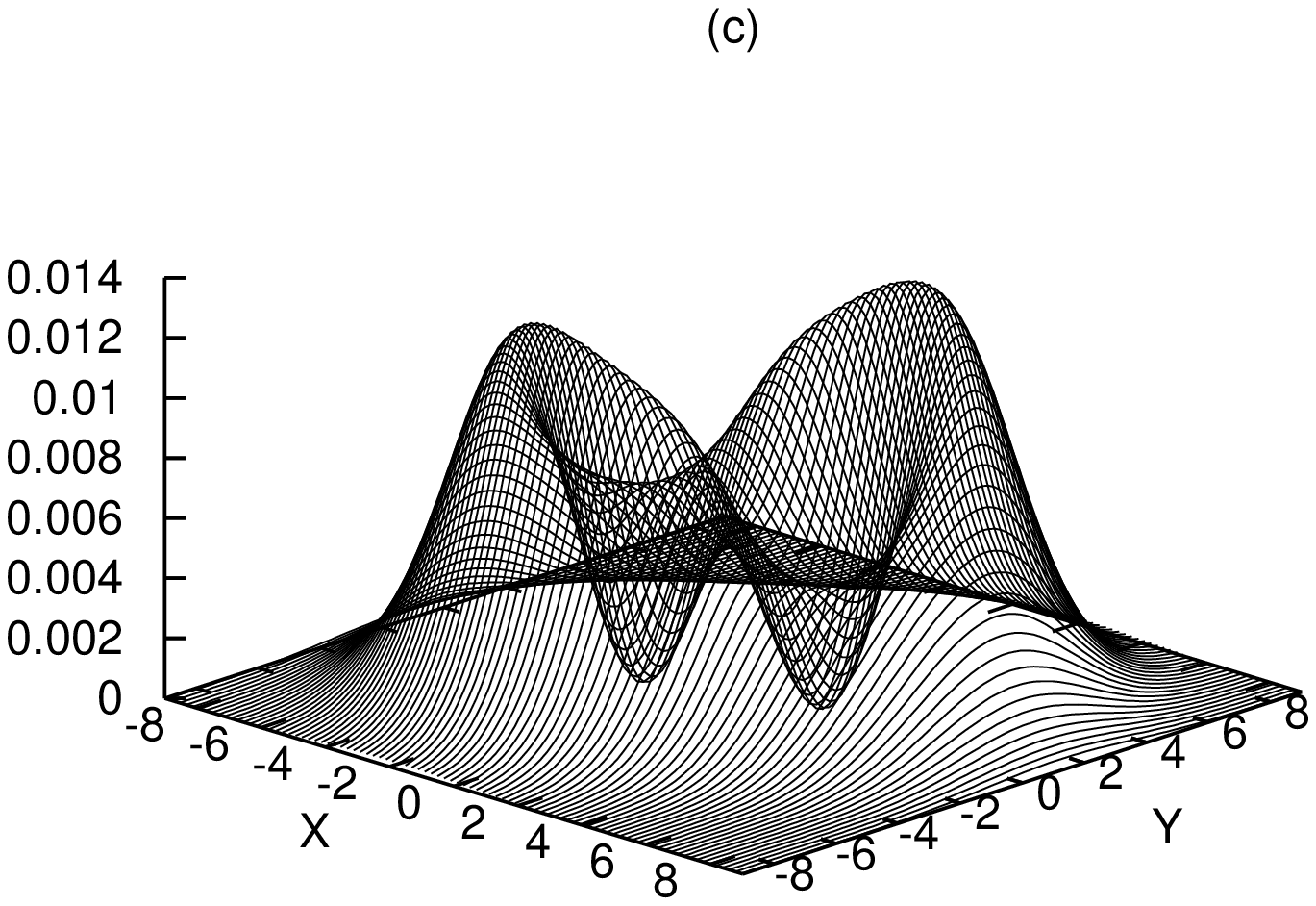} %
\includegraphics[width=6.cm,angle=-90,clip]{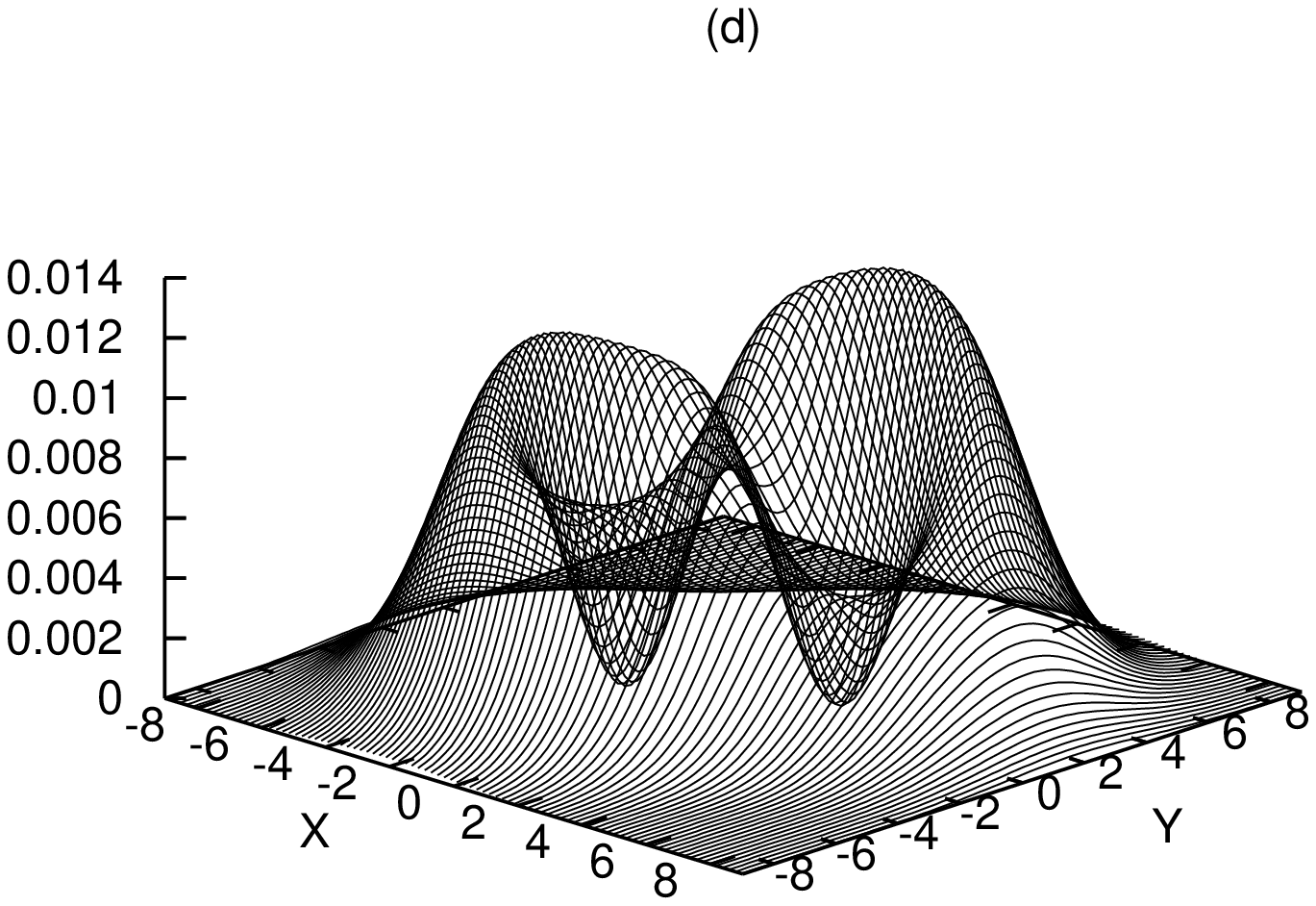}}
\caption{The same as in Fig. \protect\ref{fig7}, but for $g=-0.8$. The
snapshots pertain to $t=0$ (a), $t=44$ (b), $t=136$ (c), $t=168$ (d). The
figure displays the permanent splitting of the double vortex into a pair of
rotating unitary vortices, see the text.}
\label{fig8}
\end{figure}

At larger values of the interaction strength, $-g>0.5$, the switchings
between the states with one and two local minima is a transient regime,
which is followed by permanent splitting of the double vortex into a pair of
unitary vortices rotating around point $x=y=0$, as shown in Fig. \ref{fig8}.
Finally, at $-g\geq 1.8$ the evolution of the double vortex ends up with the
collapse.

%fig8

For the sake of the completeness of the analysis, we have also analyzed the
dynamics of the triple vortex, with $S=3$. It features a trend to the
splitting into a state with three local minima of the amplitude. At small
enough values of $|g|$, the splitting is periodically followed by the
recombination into the original triple vortex, while at larger values of $%
|g| $ it permanently splits into a set of three rotating unitary vortices.

\subsection{The periodic in-plane potential}

As discussed in Introduction, a problem of considerable interest is the
stabilization of fundamental (non-vortical) 2D solitons in the attractive
condensate loaded into a quasi-1D periodic potential, corresponding to a 1D
OL (optical lattice); the same problem has another physical realization in
terms of nonlinear optics \cite{BBB2}. In this section, we address the
problem within the framework of NPSE (\ref{npse}) with potential $%
W(x,y)=-V_{0}\cos {(2kx)}$. Accordingly, the full potential in 3D\ GPE (\ref%
{3dgpe}) is $V(x,y,z)=-V_{0}\cos {(2kx)}+z^{2}/2.$

First, the corresponding stationary solutions to the full three-dimensional
GPE, Eq. (\ref{3dgpe}), are looked for as $\Psi (\mathbf{r},t)=\psi (\mathbf{%
r})\exp (-i\mu t),$ with chemical potential $\mu $ and real function $\psi $
which obeys equation
\begin{equation}
\left[ -\frac{1}{2}\nabla ^{2}-V_{0}\cos {(2kx)}+{\frac{1}{2}}z^{2}+2\pi
g|\psi |^{2}\right] \psi =\mu \psi .  \label{psi}
\end{equation}%
This equation can be obtained by minimizing the energy functional,
\begin{equation}
E=\int d^{3}\mathbf{r}\ \psi ^{\ast }\left[ -{\frac{1}{2}}\nabla
^{2}-V_{0}\cos {(2kx)}+{\frac{1}{2}}z^{2}\psi +\pi g|\psi |\right] ^{2}\psi ,
\label{erg}
\end{equation}%
subject to constraint (\ref{norm}), which results in the following
expression for the chemical potential,
\begin{equation}
\mu =E+\pi g\int |\psi (\mathbf{r})|^{4}d\mathbf{r}.  \label{mu}
\end{equation}

To predict solitons by means of the VA (variational approximation), we use
the normalized 3D Gaussian ansatz \cite{BBB1},
\begin{equation}
\psi (\mathbf{r})={\frac{1}{\pi ^{3/4}(\sigma _{1}\sigma _{2}\sigma
_{3})^{1/2}}}\exp {\left\{ -{\frac{x^{2}}{2\sigma _{1}^{2}}}-{\frac{y^{2}}{%
2\sigma _{2}^{2}}}-{\frac{z^{2}}{2\sigma _{3}^{2}}}\right\} ,}
\end{equation}%
with widths $\sigma _{1}$, $\sigma _{2}$ and $\sigma _{3}$. Inserting the
ansatz into Eq. (\ref{erg}), one obtains
\begin{equation}
E={\frac{1}{4\sigma _{1}^{2}}}+{\frac{1}{4\sigma _{2}^{2}}}+{\frac{1}{%
4\sigma _{3}^{2}}}-V_{0}\exp {(-k^{2}\sigma _{1}^{2})}+{\frac{1}{4}}\sigma
_{3}^{2}+{\frac{g}{\sqrt{2\pi }}}{\frac{1}{\sigma _{1}\sigma _{2}\sigma _{3}}%
},
\end{equation}%
and Eq. (\ref{mu}) yields, in the same approximation, $\mu =E+g/\left( \sqrt{%
2\pi }\sigma _{1}\sigma _{2}\sigma _{3}\right) .$ Aiming to predict the
ground state in the framework of the VA, we look for values of $\sigma _{1}$%
, $\sigma _{2}$, and $\sigma _{3}$ that minimize the energy, imposing
conditions $\partial E/\partial {\sigma }_{1,2,3}=0$. In this way, we derive
the following variational equations,
\begin{gather}
4V_{0}k^{2}\sigma _{1}^{4}\exp {(-k^{2}\sigma _{1}^{2})}=1+{\frac{g}{\sqrt{%
2\pi }}}{\frac{\sigma _{1}}{\sigma _{2}\sigma _{3}}}  \notag \\
{\frac{1}{\sigma _{2}}}=-{\frac{g}{\sqrt{2\pi }}}{\frac{1}{\sigma _{1}\sigma
_{3}},~}\sigma _{3}^{4}=1+{\frac{g}{\sqrt{2\pi }}}{\frac{\sigma _{3}}{\sigma
_{1}\sigma _{2}}}.  \label{VA}
\end{gather}%
A solution corresponds to a true minimum of the energy provided that the
respective Hessian, $\partial ^{2}E/\partial \sigma _{i}\partial \sigma _{j}$%
, is positive definite.

Analysis of the variational equations demonstrates that they may yield the
ground-state solution if the OL is strong enough, \textit{viz}., $%
V_{0}>e^{2}k^{2}/16$. Under this condition, the solutions exists for $%
g_{c}<g<0$, where critical strength $g_{c}$ depends on $V_{0}$ and $k$.
Notice that, restricting the VA to the 2D setting, i.e., fixing $\sigma
_{3}=1$, cf. Eq. (\ref{Xi}), one will instead find stable solutions for $%
g_{c}<g<g_{s}$, with $g_{c}=-\sqrt{2\pi }$, and $g_{s}=-\sqrt{2\pi \left[
1-16V_{0}/(e^{2}k^{2})\right] }$ for $V_{0}<e^{2}k^{2}/16$, while $g_{s}=0$
for $V_{0}>e^{2}k^{2}/16$ \cite{BBB2}. The expression for $g_{s}$ predicts
the minimum absolute value of the self-attraction strength, above which the
state cannot remain localized in the periodic potential (in fact, the
vanishing of $g_{s}$ for $V_{0}>e^{2}k^{2}/1$ is an artifact of the VA \cite%
{Salerno,BBB2}; in the numerical solution reported for the 2D GPE with the
self-attractive cubic term, $g_{s}$ never vanishes \cite{BBB1,Salerno,BBB2}).

\begin{figure}[tbp]
{\includegraphics[width=8.cm,clip]{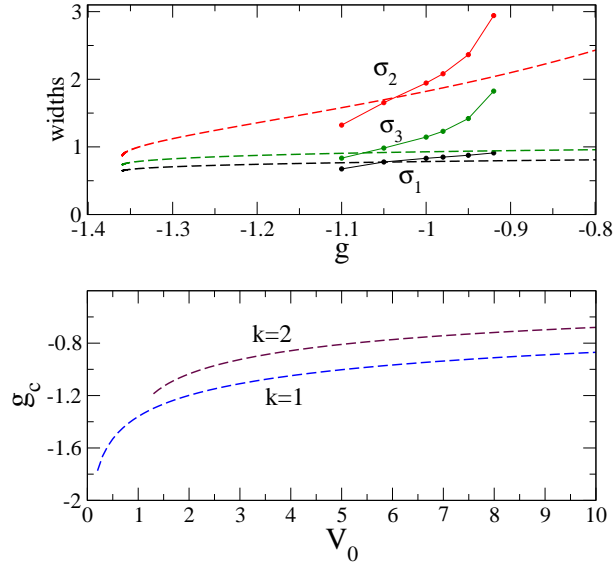}}
\caption{(Color online). In the upper panel, the dashed curves depict widths
$\protect\sigma _{1}$, $\protect\sigma _{2}$, $\protect\sigma _{3}$ of the
soliton, in the model with the quasi-1D in-plane potential, versus
self-attraction strength $g$, as predicted by the VA (variational
approximation) for $V_{0}=k=1$. The solid lines with dots show the widths as
found from a numerical solutions of the 2D NPSE. The lower panel displays
critical interaction strength $g_{c}$ for the onset of the collapse versus
amplitude $V_{0}$ of the periodic potential, as predicted by the VA, for two
different values of the wavenumber of the periodic potential, $k=1$ and $k=2$%
.}
\label{fig9}
\end{figure}

Numerically found solutions of Eqs. (\ref{VA}) are plotted versus $g$ in the
upper panel of Fig. \ref{fig9}, for $V_{0}=1$ and $k=1$ (dashed lines). We
include also the corresponding widths obtained by solving the 2D NPSE. The
lower panel of Fig. \ref{fig9} shows critical interaction strength $g_{c}$
versus amplitude $V_{0}$ of the periodic potential. At $|g|~>|g_{c}|$,
solutions for solitons do not exist anymore.

%fig9

Comparing the predictions of the VA with results produced by the 2D NPSE, we
conclude that, in the case of weak nonlinearity, the VA based on the
Gaussian ansatz is wrong, which is not surprising: in that case, solitons
are not localized like Gaussians, but rather extend over several cells of
the periodic potential \cite{Salerno,BBB2}. In particular, similar to what
was found in the 2D model with the cubic nonlinearity, and contrary to the
prediction of the VA, the above-mentioned critical value $g_{s}$, below
which (at $|g|~<\left\vert g_{c}\right\vert $) the delocalization of the 2D
soliton takes place, never drops to zero, as shown in Table 2. On the other
hand, critical strength $g_{c}$ for the onset of the collapse, as predicted
by the VA, is indeed close to what is found from the numerical solution of
the 2D NPSE, which is explained by the fact that the collapse occurs in well
localized configurations. Direct simulations of the time-dependent 2D\ NPSE
corroborate a natural expectations that the fundamental soliton is always
stable in its existence range, $g_{c}<g<g_{s}$. \bigskip

\begin{center}
\begin{tabular}{|c|c|c|}
\hline
~~~$V_0$~~~ & ~~~$g_c$~~~ & ~~~$g_s$~~~ \\ \hline
~~~$1$~~~ & ~~~$-1.12$~~~ & ~~~$-0.91$~~~ \\
~~~$2$~~~ & ~~~$-0.92$~~~ & ~~~$-0.43$~~~ \\
~~~$3$~~~ & ~~~$-0.81$~~~ & ~~~$-0.25$~~~ \\ \hline
\end{tabular}
\end{center}

\noindent Table 2. Critical strengths $g_{c}$ and $g_{s}$ between which the
fundamental soliton exists (and is stable) in the model with the quasi-1D
periodic potential with amplitude $V_{0}$ and wavenumber $k=1$, as found
from the numerical solution of the 2D NPSE. \bigskip

\begin{figure}[tbp]
{\includegraphics[width=8.cm,clip]{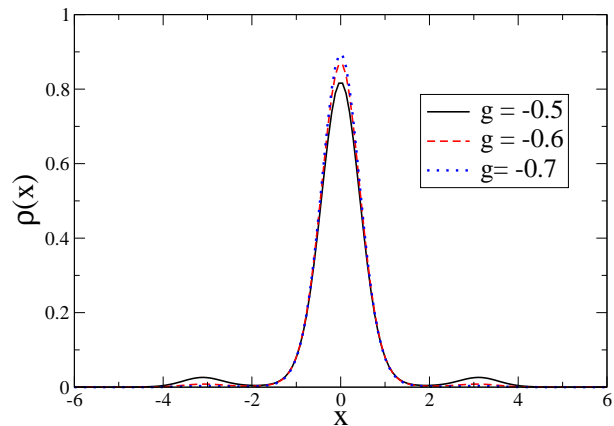}}
\caption{(Color online). Density profiles of the fundamental solitons
stabilized by the quasi-1D periodic in-plane potential are shown for three
values of self-attraction strength $g$, while the amplitude and wavenumber
of the potential are fixed, $V_{0}=-2$ and $k=1$. The density, $\protect\rho %
(x)$, is integrated in the free direction, $y$. }
\label{fig10}
\end{figure}

Finally, typical profiles of the soliton in the present version of the
model, as found from a numerical solution of Eq. (\ref{psi}), are shown in
Fig. \ref{fig10}, by means of the density integrated in the free ($y$)
direction, $\rho (x)=\int_{-\infty }^{+\infty }\psi ^{2}\left( x,y\right) dy$%
. The figure corroborates that the profile indeed features a trend to
delocalization with the decrease of $|g|$.

%fig10

\section{Conclusion}

In this work, we have reported results of the analysis of localized states
in nearly-2D (``pancake"-shaped) BEC, with both signs of the intrinsic
nonlinearity, repulsive and attractive. The localized states represent both
the ground state, with zero vorticity, and excited states in the form of
localized vortices. The main objective of the work was to apply the
nonpolynomial nonlinear Schr\"{o}dinger equation (NPSE) derived from the
full 3D GPE (Gross-Pitaevskii equation), to the analysis of the
pancake-morphed localized states. First of all, it has been demonstrated
that, in the case of the self-repulsive nonlinearity and relatively weak
parabolic (harmonic) in-plane axisymmetric trapping potential, the 2D NPSE
predicts both the radial and axial profiles of the ground states and
vortices, with the topological charge up to $S=3$, in a virtually exact
form, if compared with numerical solutions of the underlying 3D GPE. On the
contrary to that, the formal application of the ordinary 2D equation with
the cubic nonlinearity, or Thomas-Fermi approximation derived form the 3D
GPE, yields a large error. Similar conclusions have been made for the
fundamental and vortical solitons in the model with the same axisymmetric
trapping potential and self-attractive nonlinearity. As concerns the
stability, a new result was reported for the trapped vortices with $S=2$ and
$3$ in the self-attractive potential: at sufficiently small values of the
nonlinearity strength, they feature a stable dynamical regime in the form of
periodic splittings (respectively, into a pair or triplet of unitary
vortices) and recombinations.

Finally, fundamental solitons were studied in the model combining the 1D
in-plane periodic trapping potential, which does not depend on one of the
planar coordinates, and the self-attractive nonlinearity. In this case, we
compared predictions of the VA (variational approximation), derived for the
3D\ GPE, and numerical solutions of the 2D\ GPE. As a result, it has been
concluded (as might be expected) that the VA yields reasonable results,
provided that the nonlinearity is not too weak, in which case the underlying
assumption of the localization of the fundamental soliton around one cell of
the periodic potential is not valid.

\end{document}